\documentclass[12pt,a4paper]{article}
\usepackage{color, graphicx}

\def\slash#1{\mbox{$\not \!\! #1$}}

\def\lvec#1{\setbox0=\hbox{$#1$}
    \setbox1=\hbox{$\scriptstyle\leftarrow$}
    #1\kern-\wd0\smash{
    \raise\ht0\hbox{$\raise1pt\hbox{$\scriptstyle\leftarrow$}$}}
    \kern-\wd1\kern\wd0}
\def\rvec#1{\setbox0=\hbox{$#1$}
    \setbox1=\hbox{$\scriptstyle\rightarrow$}
    #1\kern-\wd0\smash{
    \raise\ht0\hbox{$\raise1pt\hbox{$\scriptstyle\rightarrow$}$}}
    \kern-\wd1\kern\wd0}

\def\diracstar#1#2{
    \setbox0=\hbox{$\gamma$}\setbox1=\hbox{$\gamma_{#1}$}
    \gamma_{#1}\kern-\wd1\kern\wd0
    \smash{\raise4.5pt\hbox{$\scriptstyle#2$}}}

\newcommand{\beq}{\begin{equation}}
\newcommand{\eeq}{\end{equation}}
\newcommand{\beqn}{\begin{eqnarray}}
\newcommand{\eeqn}{\end{eqnarray}}
\newcommand{\nn}{\nonumber}

\begin{document}
\begin{titlepage}
\pagestyle{empty}
\date{}
\title{
\bf A non-supersymmetric model with unification of electro-weak and strong interactions
\vspace*{3mm}}

\author{
        R.\ Frezzotti$^{a)}$  \quad M.\ Garofalo$^{b)}$ \quad G.C.\ Rossi$^{a)c)}$ 
}
\maketitle
\begin{center}
  {\small $^{a)}$ Dipartimento di Fisica, Universit\`a di  Roma
  ``{\it Tor Vergata}'' \\ INFN, Sezione di Roma 2}\\
    {\small Via della Ricerca Scientifica - 00133 Roma, Italy}\\
        \vspace{.2cm}
    {\small $^{b)}$ Higgs Centre for Theoretical Physics, School of Physics and Astronomy, \\
    The University of Edinburgh, Edinburgh EH9 3JZ, Scotland, UK}\\
    \vspace{.2cm}
{\small $^{c)}$ Centro Fermi - Museo Storico della Fisica\\ 
Piazza del Viminale 1 - 00184 Roma, Italy}
\end{center}

\vspace{2.cm}

\abstract{In this note we present an example of an extension of the Standard Model where unification of strong and electroweak interactions occurs at a level comparable to that of the minimal supersymmetric standard model.}
\end{titlepage}
\newpage

\section{Introduction}
\label{sec:INTRO}

\begin{figure}[htbp]   
\centerline{\includegraphics[scale=0.3,angle=0]{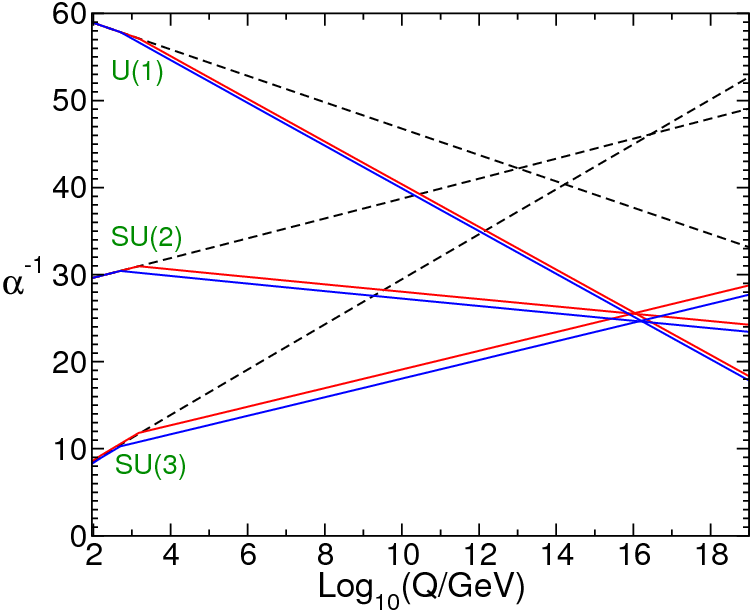}}    
\caption{\small The running of electro-weak and strong couplings in the SM (black dotted lines) and in the MSSM. The visible displacement of the blue and red curves at small scales is associated with the opening of the supersymmetry threshold taken to be either 0.5~TeV (blue curve) or 1.5~TeV (red curve) with initial conditions $\alpha_s(m_Z)=0.117$ and $\alpha_s(m_Z)=0.121$, respectively.}
\label{fig:fig1} 
\end{figure}

A desirable feature of a beyond-the-Standard-Model model (BSMM) is unification of gauge  couplings. As is well known, unification fails in the SM but it can be achieved, for instance, if the model is extended to incorporate supersymmetry~\cite{Dimopoulos:1981yj}. In fig.~\ref{fig:fig1} the 2-loop running of electro-weak and strong couplings in the SM (black dotted lines) is compared to the running in the minimal supersymmetric Standard Model (MSSM) with blue and red continuous lines referring to different supersymmetry thresholds (namely 0.5~TeV blue curve and 1.5~TeV red curve)~\cite{Martin:1997ns}.
It is undeniable that inclusion of supersymmetric partners substantially improves unification.

In this short note we want to provide an example of a model without supersymmetry where unification occurs to a similar accuracy level. The key feature of this model is that, besides the elementary particles of the SM, a new set of superstrongly interacting particles with suitably chosen hypercharge quantum numbers~\cite{WEIN} living at a scale $\Lambda_T\sim$~O(few TeV), much larger than $\Lambda_{QCD}$, is  present. 

The details of the model relevant for the considerations of this work (which we label BSMM for brevity) are specified in the Lagrangian 
\begin{eqnarray}
&& {\cal L}^{BSMM}=\frac{1}{4}\left(F^BF^B+F^WF^W+F^AF^A+F^GF^G\right)+\nn\\
&&+\sum_{f=1}^{n_g} \Big{[}\bar q^f_L  \slash D^{BWA} q^f_L+\bar q_R^{f\,u} \slash D^{BA} q^{f\,u}_R+\bar q_R^{f\,d} \slash D^{BA} q_R^{f\,d}+\nn \\
&&\qquad+\bar \ell_L^f \slash D^{BW} \ell_L^f+\bar \ell_R^{f\,u} \slash D^{B} \ell_R^{f\,u}+\bar \ell_R^{f\,d} \slash D^{B} \ell_R^{f\,d}\Big{]}+\nn\\
&&+\sum_{s=1}^{\nu_Q} \Big{[}\bar Q^s_L \slash D^{BWAG} Q^s_L+\bar Q_R^{s\,u} \slash D^{BAG} Q_R^{s\,u} +\bar Q_R^{s\,d}  \slash D^{BAG} Q_R^{s\,d}\Big{]}+\nn\\
&&+\sum_{t=1}^{\nu_L} \Big{[}\bar L_L^t \slash D^{BWG} L_L^t+\bar L_R^{t\,u} \slash D^{BG} L_R^{t\,u}+\bar L_R^{t\,d} \slash D^{BG} L_R^{t\,d} \Big{]}+\nn\\
&&+\frac{1}{2}\mbox{Tr}\left[D_\mu^{BW}\Phi^\dagger D_\mu^{BW}\Phi \right]+\frac{\mu_0^2}{2}\mbox{Tr}\left[\Phi^\dagger\Phi \right]+\frac{\lambda_0}{4} \left(\mbox{Tr}\left[\Phi^\dagger\Phi \right]\right)^2\, ,
\label{LMOD}
\end{eqnarray}
where the new set of superstrongly interacting particles (SIPs), including $Q$, $L$, and superstrong gauge bosons, $G$, is gauge invariantly coupled to SM gauge bosons ($B,W,A$) and fermions ($q, \ell$).

We have indicated with $D_\mu^X$ the covariant derivative with respect to the group transformations of which $\{X\}$ are the associated gauge bosons. The most general expression of the covariant derivative is
\begin{equation}
D_\mu^{BWAG}=\partial_\mu-iYg_YB_\mu-ig_w\tau^r W^r_\mu-ig_s\frac{\lambda^a}{2} A^a_\mu-ig_T\frac{\lambda^{\alpha}_T}{2} G^{\alpha}_\mu \, , \label{DER}
\end{equation}
where $Y, \tau^r (r=1,2,3), \lambda^a (a=1,2, \ldots, N_c^2-1)$ and $\lambda^{\alpha}_T (\alpha=1,2,\ldots, N_T^2-1)$ are, respectively, the U$_Y(1)$ hypercharge and the generators of the SU$_L$(2), SU($N_c=3$), SU($N_T=3$) group with $g_Y, g_w, g_s, g_T$ denoting the corresponding gauge couplings~\footnote{We use the notation $g_w$ for the SU$_L$(2) gauge coupling. This can be a little confusing as the latter is usually denoted by $g$ in standard textbooks~\cite{WEIN}.}. We notice that $Q$ are subjected to electro-weak, strong and superstrong interactions, while $L$ to electro-weak and superstrong interactions only.

For the SU$_L$(2) SM matter doublets we use the notation $q_L=(u_L,d_L)^T$ and $\ell_L=(\nu_L,e_L)^T$. Right-handed components are SU$_L$(2) singlets and are denoted by $q^u_R$, $q^d_R$ and $\ell^u_R$, $\ell^d_R$. A similar notation is used for $Q$ and $L$ fermions. Some of the formulae below will be given for the general case of $n_g$ SM families and $\nu_Q, \nu_L$ generations of SIPs. 

The scalar field, $\Phi$, is a $2\times2$ matrix with $\Phi=(\phi,-i\tau^2 \phi^*)$ and $\phi$ an iso-doublet of complex scalar fields, that feels U$_Y(1)$ and SU$_L$(2), but not SU($N_c=3$) and SU($N_T=3$), gauge interactions. 

Concerning mass terms, for the $\beta$-function calculation of interest in this paper we only need to say that SIPs have masses O($\Lambda_T\gg \Lambda_{QCD}$) with $\Lambda_T$ a scale in the few TeV range. 

The motivation for considering the model~(\ref{LMOD})~\cite{Frezzotti:2014wja} (see also ref.~\cite{Frezzotti:2013raa}) is outlined in sect.~\ref{sec:MODEL}.

\section{1-loop $\beta$-functions}

In this section we present the relevant formulae for the evaluation of the 1-loop running of the four gauge couplings, $g_Y, g_w, g_s, g_T$ associated to the U$_Y(1)$, SU$_L$(2), SU($N_c=3$) and SU($N_T=3$) gauge groups, respectively.

As we shall see, a special role in achieving unification is played by the hypercharge assignment of SIPs. Indeed, as worked out in the Weinberg book~\cite{WEIN}, there exist two possible solutions to the anomaly cancellation equations as far as hypercharge assignment is concerned. Besides the standard assignment (that we recall in Table~\ref{tab:standard}) in which U(1) anomalies are cancelled between quarks and leptons, there is another solution in which anomalies are cancelled within quark and lepton sectors separately. They are reported in Table~\ref{tab:alternative} where we display the simplest choice consistent with the assumption that right-handed particles are SU$_L(2)$ singlets and ${\cal Q}=T_3+Y$. Table~\ref{tab:alternative} corresponds to taking $|{\cal Q}_Q|=|{\cal Q}_L|=1/2$.
\begin{table}[htbp]
\begin{center}
\begin{tabular}{|l|l|}
\hline 
\vspace{.1cm}
$q$ & $\ell$   \\
\hline
\hline
\vspace{.1cm}
 $y_{u_L}=\frac{2}{3}-\frac{1}{2}=\frac{1}{6}$ & $y_{\nu_L}=0-\frac{1}{2}=-\frac{1}{2}$
\\
\hline
 \vspace{.1cm}  
 $y_{u_R}=\frac{2}{3}-0=\frac{2}{3}$ & $y_{\nu_R}=0$ \\
\hline
 \vspace{.1cm}  
 $y_{d_L}=-\frac{1}{3}+\frac{1}{2}=\frac{1}{6}$ & $y_{el_L}=-1+\frac{1}{2}=-\frac{1}{2}$   \\
\hline
 \vspace{.1cm}  
 $y_{d_R}=-\frac{1}{3}-0=-\frac{1}{3}$ & $y_{el_R}=-1-0=-1$   \\
\hline
\hline
 \vspace{.1cm}  
$\sum y^2_q =\frac{22}{36}$ & $\sum y^2_\ell =\frac{3}{2}$\\
\hline
\end{tabular}
\caption{Hypercharges of SM fermions}
\label{tab:standard}
\end{center}
\end{table}
\begin{table}[htbp]
\begin{center}
\begin{tabular}{|l|l|}
\hline
\vspace{.1cm}
$Q$  & $L$ 
 \\
\hline
\hline
\vspace{.1cm}
$y_{U_L}=\frac{1}{2}-\frac{1}{2}=0$ & $y_{N_L}=\frac{1}{2}-\frac{1}{2}=0$   \\
\hline
\vspace{.1cm}
$y_{U_R}=\frac{1}{2}-0=\frac{1}{2}$ & $y_{N_R}=\frac{1}{2}-0=\frac{1}{2}$   \\
\hline
\vspace{.1cm}
$y_{D_L}=-\frac{1}{2}+\frac{1}{2}=0$ & $y_{L_L}=-\frac{1}{2}+\frac{1}{2}=0$ \\
\hline
\vspace{.1cm}
$y_{D_R}=-\frac{1}{2}-0=-\frac{1}{2}$ & $y_{L_R}=-\frac{1}{2}-0=-\frac{1}{2}$ \\
\hline
\hline
\vspace{.1cm}
$\sum y^2_Q =\frac{1}{2}$ & $\sum y^2_L =\frac{1}{2}$ \\
\hline
\end{tabular}
\caption{Non standard hypercharge assignments.}
\label{tab:alternative}
\end{center}
\end{table}

\subsection{1-loop $\beta$-functions of the BSMM~(\ref{LMOD})}

With the standard definitions 
\beq
\beta_x(g_x)=\mu\frac{dg_x}{d\mu}\, , \qquad x=T,s,w,Y 
\label{BETAZ}
\eeq
and taking the assignment of Table~\ref{tab:alternative} for the SIP hypercharges, one gets
\beqn
\hspace{-1.2cm}&&\beta_{T}^{BSMM}=-\left[\frac{11}{3}N_T-\frac{4}{3}(N_c\nu_Q+\nu_L)\right]\frac{g_T^3}{(4\pi)^2}\, ,\nn\\
\hspace{-1.2cm}&&\beta_{s}^{BSMM}=-\left[ \frac{11}{3}N_c-\frac{4}{3}(N_T\nu_Q+n_g) \right]\frac{g_s^3}{(4\pi)^2}\, , \nn\\
\hspace{-1.2cm}&& \beta_{w}^{BSMM}=-\left[ 2 \frac{11}{3}-\frac{1}{3}n_g(N_c+1)-\frac{1}{3} N_T(N_c\nu_Q+\nu_L)-\frac{1}{6} \right]\frac{g_w^3}{(4\pi)^2}\,, \nn\\
\hspace{-1.2cm}&& \beta_{Y}^{BSMM}=\left\{\frac{2}{3}\left[\left(\frac{22}{36}N_c+\frac{3}{2} \right)n_g+\frac{1}{2} N_T(N_c\nu_Q+\nu_L)\right]+\frac{1}{6}\right\}\frac{g_Y^3}{(4\pi)^2} \, ,\label{OLBSMM}
\eeqn
where for generality we have left unspecified the rank of the strong and superstrong gauge groups ($N_c$ and $N_T$), the number of SM families ($n_g$) and the number of SIPs generations ($\nu_Q$ and $\nu_L$).

If, instead, also for SIPs the standard hypercharge assignment is taken, only $\beta_{Y}$ is modified and becomes   
\begin{equation}
\beta^{BSMM}_{Y\,st}\!=\left\{\frac{2}{3}\left[\left(\frac{22}{36}N_c+\frac{3}{2} \right)n_g+\left(\frac{22}{36}N_c\nu_Q +\frac{3}{2}\nu_L \right)\!N_T\right] \!+\!\frac{1}{6}\right\}\frac{g_Y^3}{(4\pi)^2} \, .\label{OLBSMMSH} 
\end{equation}

\subsection{1-loop SM $\beta$ function}
For comparison we report the 1-loop $\beta$-functions of the SM~\cite{Machacek:1983tz} that read
\beqn
&&\beta_{s}^{SM}=-\left(\frac{11}{3}N_c-\frac{4}{3} n_g\right)\frac{g_s^3}{(4\pi)^2}\, , \nn\\
&&\beta_{w}^{SM}=-\left[2\frac{11}{3}-\frac{1}{3}n_g(N_c+1)-\frac{1}{6}\right]\frac{g_w^3}{(4\pi)^2}\, ,\nn\\
&& \beta_{Y}^{SM}=\left[\frac{2}{3}\left(\frac{22}{36}N_c+\frac{3}{2} \right)n_g+\frac{1}{6} \right]\frac{g_Y^3}{(4\pi)^2}\, .\label{OLSM}
\eeqn 

\section{GUT normalization}

In order to check whether or not on the basis of the running implied by the above equations there is (an even approximate) unification, one has to determine the normalization of the couplings that should unify by requiring that the generators of the U$_Y(1)$, SU$_L$(2), SU($N_c=3$) and SU($N_T=3$) groups are among the generators of the allegedly existing simple compact unification group, G$_{GUT}$.
 
\subsection{BSMM}

For the BSMM of eq.~(\ref{LMOD}) the GUT normalization condition reads
\begin{equation}
\mbox{Tr}\left[ (g_YY)^2\right]=\mbox{Tr}\left[ (\frac{1}{2}g_w\tau^3)^2\right]=\mbox{Tr}\left[ (\frac{1}{2}g_s\lambda^3)^2\right]=\mbox{Tr}\left[ (\frac{1}{2}g_T\lambda^{3}_{T})^2\right] \, ,\label{CGUT}
\end{equation}
where the sum in the trace is extended over all the fermions building up the putative irreducible representation of the GUT group and it is normalized so that each Weyl component contributes one unit. With the alternative hypercharge assignment of Table~\ref{tab:alternative} one gets in this way 
\beqn
 \hspace{-.8cm}&&\mbox{Tr}\left[ (g_YY)^2\right]= \nn\\
 \hspace{-.8cm}&&\quad=\Big{[}2n_g\left(\frac{1}{2}\right)^2\!+n_g(-1)^2\!+2n_gN_c\left(-\frac{1}{6}\right)^2\!+n_gN_c\left(\frac{2}{3}\right)^2\!+n_gN_c\left(-\frac{1}{3}\right)^2+\nn\\
\hspace{-.8cm}&&\quad+N_T\left(\frac{1}{4}+\frac{1}{4}\right)(\nu_L+\nu_QN_c)\Big{]}g_Y^2=\nn\\
\hspace{-.8cm}&&\quad=\left[ n_g\left(\frac{3}{2}+N_c\frac{22}{36}\right)+\frac{N_T}{2}(\nu_L+\nu_QN_c)\right]g_Y^2\, ,\nn\\
\hspace{-.8cm}&&\mbox{Tr}\left[ (\frac{1}{2}g_w\tau^3)^2\right]=\left[ n_g[\frac{1}{2}(N_c+1)]+N_T[\frac{1}{2}(\nu_QN_c+\nu_L)]\right]g_w^2\, , \nn\\
\hspace{-.8cm}&&\mbox{Tr}\left[ (\frac{1}{2}g_s\lambda^3)^2\right]= 2(n_g+N_T\nu_Q)g_s^2\, ,\nn\\
\hspace{-.8cm}&&\mbox{Tr}\left[\left(\frac{1}{2}g_T\lambda^{3}_T\right)^2 \right]=2(\nu_L+\nu_QN_c) g_T^2\, .\label{NORBSMM}
\eeqn
Setting $N_c=N_T=n_g=3$ and $\nu_L=\nu_Q=1$ in eqs.~({\ref{NORBSMM}}), one concludes that, up to an (irrelevant) overall constant, the couplings that we need to consider in order to study unification are 
\beqn
&& g_1^2:=\frac{4}{3}\,g_Y^2\, , \qquad g_2^2:=g^2_w\, ,\qquad  g_3^2:=g_s^2\, ,\qquad g_4^2:=\frac{2}{3}\,g_T^2\, .
\label{NORGUTBSMM}
\eeqn
For the BSMM with the standard hypercharge assignment of Table~\ref{tab:standard} one finds
\beqn
\hspace{-.8cm}&&\mbox{Tr}\left[ (g_Y Y_{\,st})^2\right]=\nn\\
\hspace{-.8cm}&&\quad=\Big{[}2n_g\left(\frac{1}{2}\right)^2+n_g(-1)^2+2N_cn_g\left(-\frac{1}{6}\right)^2+N_cn_g\left(\frac{2}{3}\right)^2+N_cn_g\left(-\frac{1}{3}\right)^2+ \nn\\
\hspace{-.8cm}&&\quad+2N_T\nu_L\left(\frac{1}{2}\right)^2+\nu_LN_T(-1)^2\!+2N_c \nu_Q N_T\left(-\frac{1}{6}\right)^2+N_c \nu_Q N_T\left(\frac{2}{3}\right)^2+\nn\\
\hspace{-.8cm}&&\quad+N_c\nu_Q N_T\left(-\frac{1}{3}\right)^2\Big{]}g_Y^2\, , \label{EQ1}
\eeqn
so that with $N_c=N_T=n_g=3$ and $\nu_L=\nu_Q=1$ the set of couplings specified in~(\ref{NORGUTBSMM}) should be replaced by 
\begin{equation}
g_1^2:=\frac{5}{3}\,g_Y^2\, , \qquad g_2^2:=g^2_w\, ,\qquad  g_3^2:=g_s^2\, ,\qquad g_4^2:=\frac{2}{3}\,g_T^2\, .
\label{NORGUTBSMMSH}
\end{equation}

\subsection{SM}
The analogous normalization formulae for the SM gauge couplings unification (for $N_c=3$) read
\begin{equation}
\mbox{Tr}\left[ (g_YY)^2\right]=\mbox{Tr}\left[ (\frac{1}{2}g_w \tau^3)^2\right]= \mbox{Tr}\left[ (\frac{1}{2}g_s\lambda^3)^2\right] \, ,
\label{tr_GUT_SM}
\end{equation}
with
\beqn
\hspace{-.8cm}&&\mbox{Tr}\left[ (g_YY)^2\right]\!=\!\left[2n_g \!\left(\frac{1}{2}\right)^2\!\!+\!n_g(-1)^2\!+\!6n_g\!\left(\!-\frac{1}{6}\right)^2\!\!+\! 3n_g\left(\frac{2}{3}\right)^2\!\!+\!3n_g\left(\!-\frac{1}{3}\right)^2\right]g_Y^2\! =\nn\\
\hspace{-.8cm} &&\quad=\frac{10}{3}n_gg_Y^2 \, , \nn\\
\hspace{-.8cm}&&\mbox{Tr}\left[ (\frac{1}{2}g_w \tau^3)^2\right]=(3n_g+n_g)\left[\left(\frac{1}{2}\right)^2+\left(\frac{1}{2}\right)^2\right]g_w^2=2n_gg_w^2\, , \nn\\
\hspace{-.8cm}&&\mbox{Tr}\left[ (\frac{1}{2}g_s\lambda^3)^2\right]=\left[4n_g\left(\frac{1}{2}\right)^2+4n_g\left(-\frac{1}{2}\right)^2\right]g_s^2=2n_gg_s^2\, ,
\label{NORMSM}
\eeqn
from which for $n_g=3$ one gets
\begin{equation}
g_1^2:=  \frac{5}{3}g_Y^2\, ,\qquad g_2^2:=g_w^2\, ,\qquad g_3^2:=g_s^2\, .
\label{NORGUTSM}
\end{equation}

\subsection{Results for $N_c=3$, $n_g=3$, $N_T=3$, $\nu_Q=1$, $\nu_L=1$}

Putting together the formulae for the 1-loop beta functions (eqs.~(\ref{OLBSMM}), (\ref{OLBSMMSH}) and~(\ref{OLSM})) with the corresponding normalizations (eqs.~(\ref{NORGUTBSMM}), (\ref{NORGUTBSMMSH}) and~(\ref{NORGUTSM})), one gets the RG equations 
\beq
\frac{d g_i}{d\log \mu}=\beta_{g_i}\, , \quad i=1,2,3,4 \, .
\label{RUNNING}
\eeq
The $\beta_{g_i}$ appropriate for the various cases we have considered are given in the next subsections.

\subsubsection{BSMM 1-loop $\beta$-functions with GUT normalization}

\begin{eqnarray}
&& \beta_{g_1}=\frac{65}{8}\frac{g^3_1}{(4\pi)^2}\, ,\nn\\
&& \beta_{g_2}=\frac{5}{6}\frac{g^3_2}{(4\pi)^2}\, ,\nn\\
&& \beta_{g_3}=-3\frac{g^3_3}{(4\pi)^2}\, ,\nn\\
&& \beta_{g_4}=-\frac{17}{2}\frac{g^3_4}{(4\pi)^2}\, . 
\label{BBSMM}
\end{eqnarray}
If one were to use the standard choice of hypercharges for $Q$ and $L$ particles of Table~\ref{tab:standard}, $\beta_{g_1}$ must be replaced by
\begin{eqnarray}
\beta_{g_1}^{Y\,st}=\frac{81}{10}\frac{g^3_1}{(4\pi)^2}\,.
\end{eqnarray}

\subsubsection{SM 1-loop $\beta$-functions with GUT normalization}

\begin{eqnarray}
&& \beta_{g_1}=\frac{41}{10} \frac{g_1^3}{(4\pi)^2} \, ,\nn\\
&& \beta_{g_2}=-\frac{19}{6} \frac{g_2^3}{(4\pi)^2}\, ,\nn\\
&& \beta_{g_3}=-7 \frac{g_3^3}{(4\pi)^2} \label{beta_1(g_3)SM}\, .
\end{eqnarray}
One notices the change of sign of the $\beta_{g_2}$ coefficient in the BSMM~(\ref{LMOD}) with respect to the case of the SM.

\section{Unification of couplings}
\label{sec:PLOT}

\begin{figure}[htbp]   
\centerline{\includegraphics[scale=0.8,angle=0]{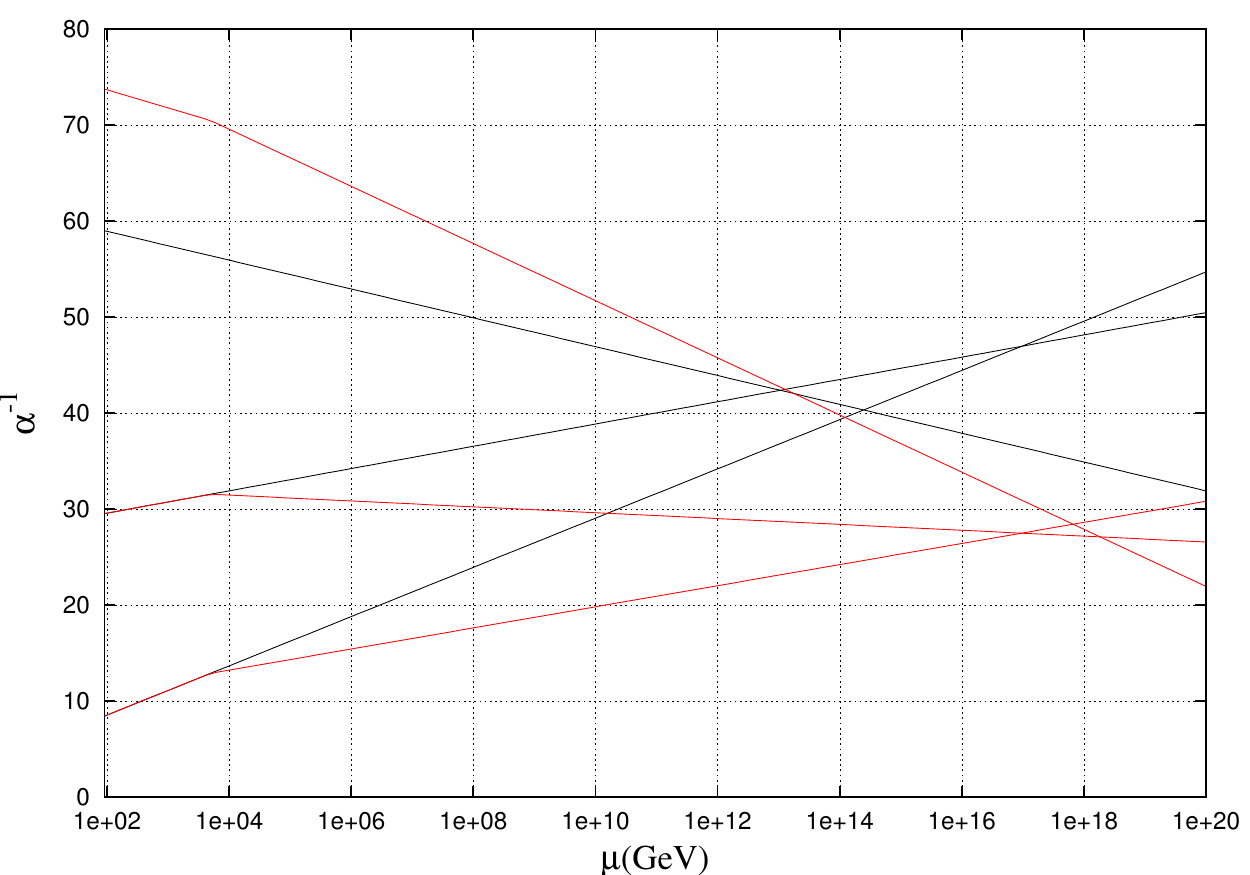}}    
\caption{\small The 1-loop running of electro-weak and strong couplings in the BSMM (red curves) and in the SM (black curves). The visible change of slope of red curves at ``low'' scales is associated with the opening of the superstrong threshold that we take to be 5~TeV.}
\label{fig:fig4} 
\end{figure}

In fig.~\ref{fig:fig4} we plot the 1-loop running of the U$_Y(1)$, SU$_L$(2) and SU($N_c=3$) inverse square gauge coupling, $\alpha_i^{-1}=4\pi/g^2_i$, in the BSMM of eq.~(\ref{LMOD}) with the choice of the hypercharges reported in Table~\ref{tab:alternative} (red curves), compared to the running in the SM (black curves). The input values of the inverse gauge couplings at low energy (i.e.\ at $m_Z\sim 91$~GeV) for the SM have been fixed by taking the recent PDG data~\cite{PDG}
\beqn
 && \alpha_1^{-1}(m_Z)=59.01\pm 0.02 \, ,\nn\\
 && \alpha_2^{-1}(m_Z)=29.57\pm 0.02\, ,\nn \\
 && \alpha_3^{-1}(m_Z)=8.45\pm 0.05 \, .\label{INIT}
\eeqn

\noindent Because of the different normalization of the U$_Y(1)$ coupling (compare eqs.~(\ref{NORGUTBSMM}) and~(\ref{NORGUTBSMMSH})), the $\alpha_1^{-1}$ input values  of the red (BSMM) and black (SM) curve are different. For the BSMM the input value of $\alpha_1^{-1}$ must be taken to be  
\begin{equation}
\alpha_1^{-1}(m_Z)=73.76\pm0.02\, .
\end{equation}

The key result of the present investigation is the observation that in the model~(\ref{LMOD}) unification occurs to a much better level than in the SM. It is worth noticing the non-negligible effect due to the opening of the superstrongly interacting degrees of freedom threshold (that for definiteness we have set at 5~TeV). 

\begin{figure}[htbp]   
\centerline{\includegraphics[scale=0.8,angle=0]{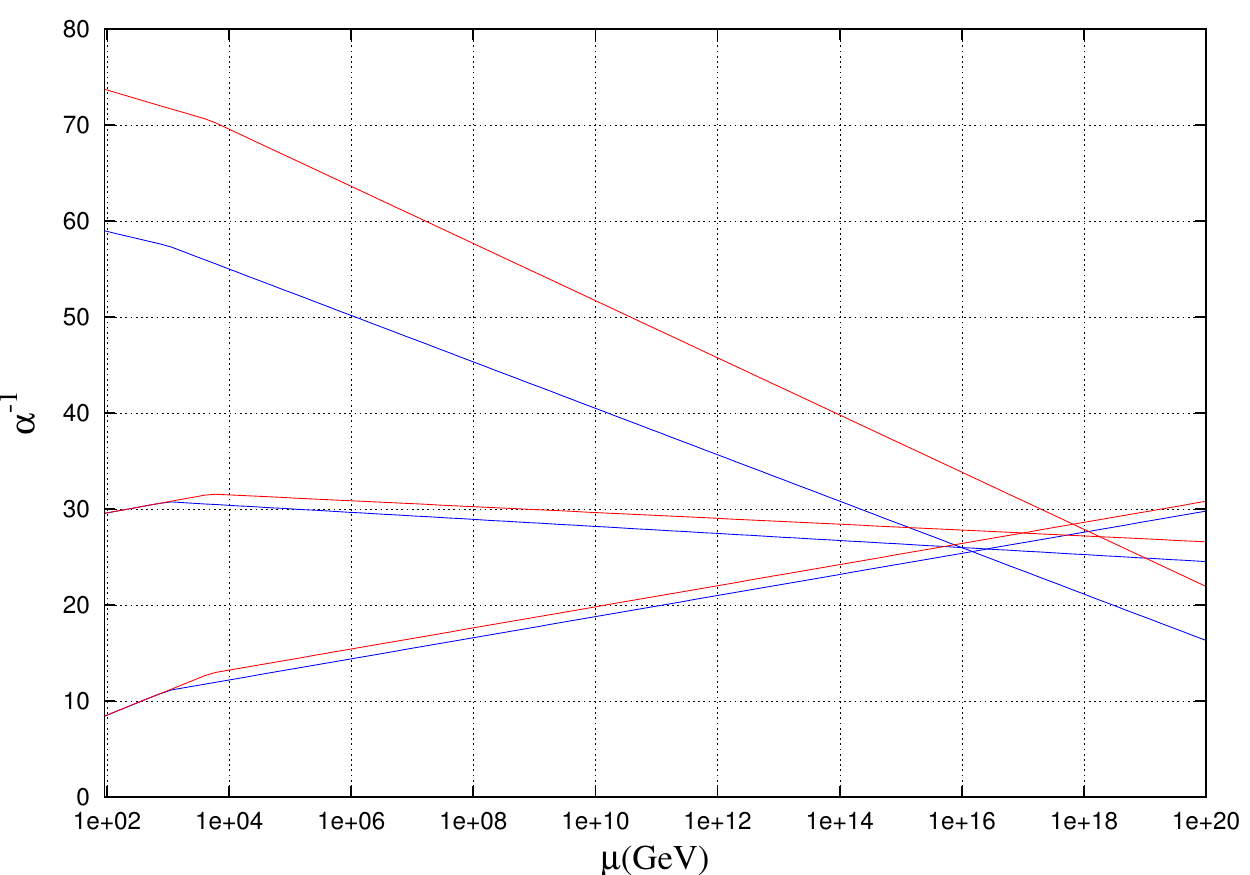}}    
\caption{\small The 1-loop running of electro-weak and strong couplings in the BSMM (red curves) and in the MSSM (blue curves). The supersymmetry and superstrong thresholds have been set at $\Lambda_{T}=5$ and $\Lambda_{MSSM}=1$~TeV, respectively.}
\label{fig:fig5} 
\end{figure}

To appreciate the quality of unification one may compare the BSMM running with that of the MSSM controlled by the 1-loop $\beta$-functions~\cite{Martin:1997ns}.
\beqn
&&\beta_{g_1}^{MSSM}=\frac{33}{5} \frac{g_1^3}{(4\pi)^2}\, ,\nn\\
&&\beta_{g_2}^{MSSM}=\frac{g_2^3}{(4\pi)^2}\, ,\nn\\
&&\beta_{g_3}^{MSSM}=-3 \frac{g_3^3}{(4\pi)^2}\, ,
\label{BETAMSSM}
\eeqn
where now $g_1^2=5/3\,g^2_Y$. The comparison in shown in fig.~\ref{fig:fig5}. The blue curves refer to the 1-loop running in the MSSM, employing the input values specified in eqs.~(\ref{INIT}). The supersymmetry and superstrong thresholds have been set at unequal values (namely $\Lambda_{MSSM}=1$ and $\Lambda_{T}=5$~TeV, respectively). Notice, in fact, that the curves representing  $\alpha_3^{-1}$ only differ because different values for the opening of the supersymmetry and superstrong thresholds have been taken, while in the case of  $\alpha_2^{-1}$ also the 1-loop coefficients of the $\beta$-function are different (second line of eqs.~(\ref{BETAMSSM}) and~(\ref{BBSMM}), respectively). The evolution of $\alpha_1^{-1}$ in the two models differ both because of the input values resulting from the different GUT normalization of the U$_Y(1)$ generator and of the unequal 1-loop coefficients of $\beta_{g_1}$. We stress that the quality of unification slightly depends on the exact magnitude of the threshold values.

\begin{figure}[htbp]   
\centerline{\includegraphics[scale=0.6,angle=0]{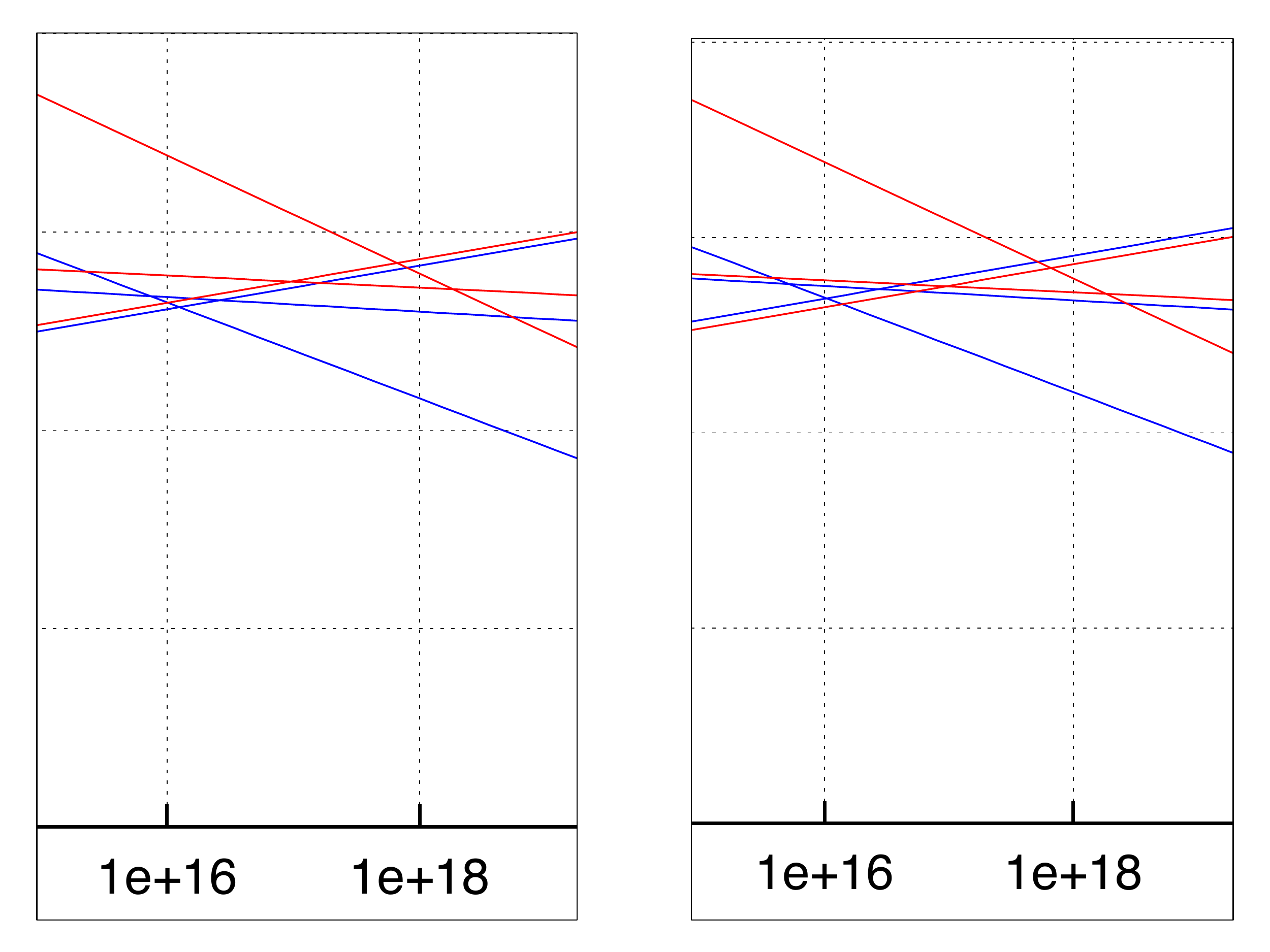}}    
\caption{\small The 1-loop running of electro-weak and strong couplings in the BSMM (red curves) and in the MSSM (blue curves). The supersymmetry thresholds have been set at $\Lambda_{MSSM}=3$ (left panel) and $\Lambda_{MSSM}=10$~TeV (right panel). The scale $\Lambda_T$ relevant for the BSMM curves is kept at 5~TeV. The dashed grid is the same as in fig.~\ref{fig:fig5}.}
\label{fig:fig6} 
\end{figure}

In our opinion the level of unification provided by the particle content of the Lagrangian~(\ref{LMOD}) compares rather well with what one gets in the MSSM, especially if one notices that an increase of the SUSY scale (for instance going from the values of Fig.~\ref{fig:fig1} to 3 or even 10~TeV as it might be necessary in the light of the recent LHC data) tends to worsen the level of coupling unification. In the two panels of fig.~\ref{fig:fig6} we show a blow up of the coupling crossing region corresponding to $\Lambda_{MSSM}=3$ and 10~TeV. The scale $\Lambda_T$ is kept at 5~TeV. We see that the blue triangle gets wider as $\Lambda_{MSSM}$ is increased.

The uncertainties on unification coming from SUSY thresholds is numerically similar to the 2-loop running corrections in MSSM~\cite{Martin:1997ns} and the order of magnitude of the ``unknown'' GUT threshold effects. Altogether all these uncertainties make difficult to predict the value of the unified inverse coupling with an absolute accuracy better than one unit.

\subsection{Observations}
\label{sec:OBS}

We conclude this section with a few observations.

\subsubsection{Hypercharge assignments}
\label{sec:OHA}

\begin{figure}[htbp]   
\centerline{\includegraphics[scale=0.8,angle=0]{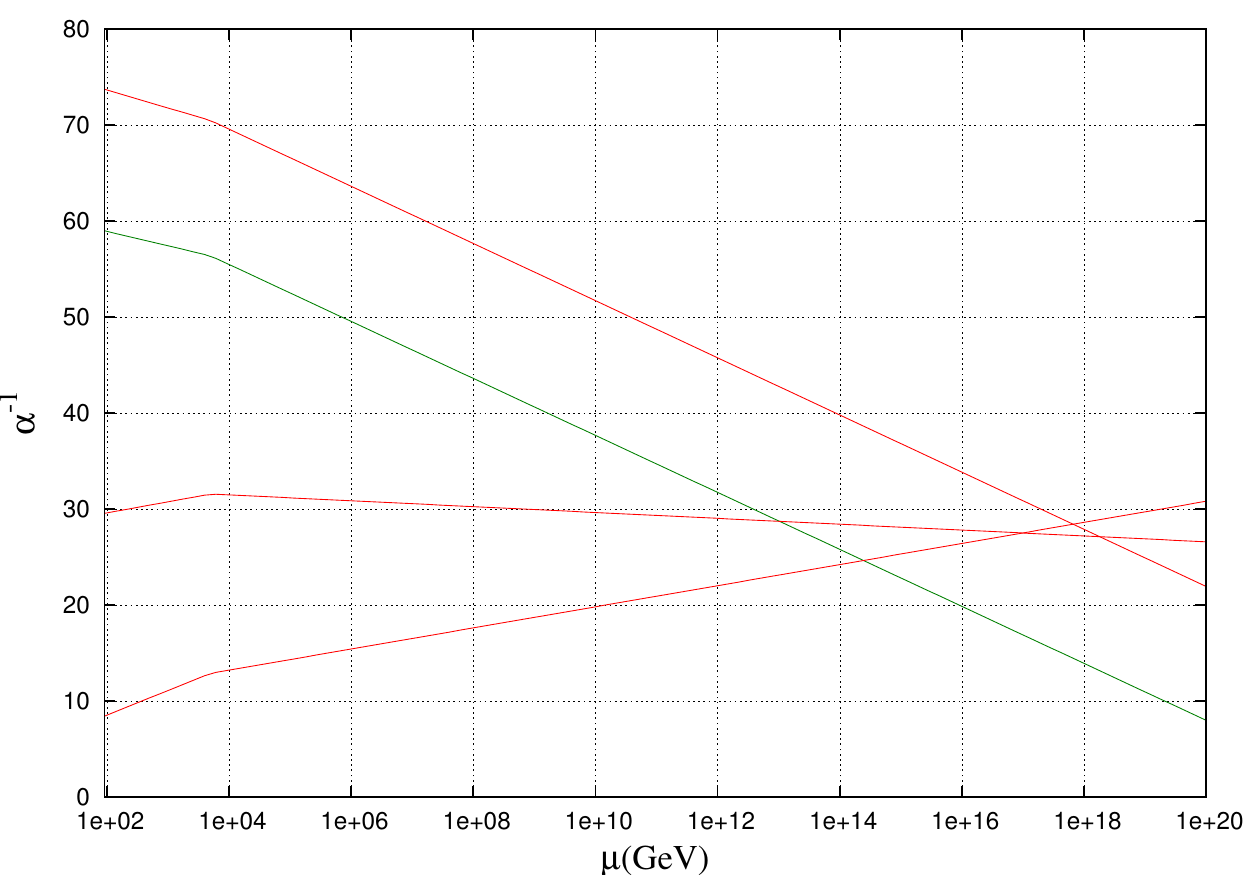}}    
\caption{\small The 1-loop running of electro-weak and strong couplings in the BSMM with the hyperchage assignment of Table~\ref{tab:alternative} (red curves) and with the hyperchage assignment of Table~\ref{tab:standard} (green curve).}
\label{fig:fig8} 
\end{figure}

The hypercharge assignment in Table~\ref{tab:alternative} is the one for which Weinberg~\cite{WEIN} writes that it ``resembles nothing observed in nature''. The reason for taking it for the hypercharges of SIPs is that with the more standard choice of Table~\ref{tab:standard} one would not get a unification as good as the one visible in fig.~\ref{fig:fig5}. The difference is evident in fig.~\ref{fig:fig8} where we compare the runnings that in the BSMM are produced with the two types of hypercharge assignments. Clearly no unification can be achieved if the green curve, corresponding to the Table~\ref{tab:standard} assignment, describes the one-loop running of $\alpha_1^{-1}$.

It must be noticed that the hypercharge assignment of Table~\ref{tab:alternative} yields $Q$ and $L$ elementary particles with electric charges $\pm e/2$, hence superstrongly confined ``hadrons'' have electrical charge quantized in units of $e/2$.

\subsubsection{2-loop $\beta$-functions \& threshold effects}
\label{sec:TOHA}

We have extended the calculations of all the $\beta$-functions to 2-loops~\cite{thesis}. We do not report the corresponding (rather cumbersome) formulae here because 2-loop terms do not modify in any essential way the previous plots, hence the quality of unification of gauge couplings visible in fig.~\ref{fig:fig4}. We recall that at 2-loops the RG equations become much more involved as the RG evolution of each coupling depends on all the others. 

As we do not know what the full UV completion of the fundamental theory~(\ref{LMOD}) could be, and consistently with our decision of (momentarily) neglecting 2-loop terms, we refrain from giving estimates of possible effects due threshold opening of GUT degrees of freedom around the GUT scale.

In any case 2-loop corrections and threshold effects tend to be of comparable numerical magnitude and may affect the values of the inverse coupling at (around) the unification scale at the level of about one unit.

\subsubsection{Unification with superstrong interactions}
\label{sec:SSU}

A very indirect clue on the UV structure of the GUT theory can come from the interesting observation that unification of {\it all the  four} gauge couplings [U$_Y(1)$, SU$_L$(2), SU($N_c=3$) and SU($N_T=3$)] can be achieved if a certain number, $N_S$, of purely SIPs are included in the model~(\ref{LMOD}) with a Lagrangian of the form $\sum_{h=1}^{N_S}\left(\bar\psi^h \slash D^G \psi^h+m_h\bar\psi^h\psi^h\right)$, where $m_h$ is an O($\Lambda_{GUT}$) mass scale. 

\begin{figure}[htbp]   
\centerline{\includegraphics[scale=0.8,angle=0]{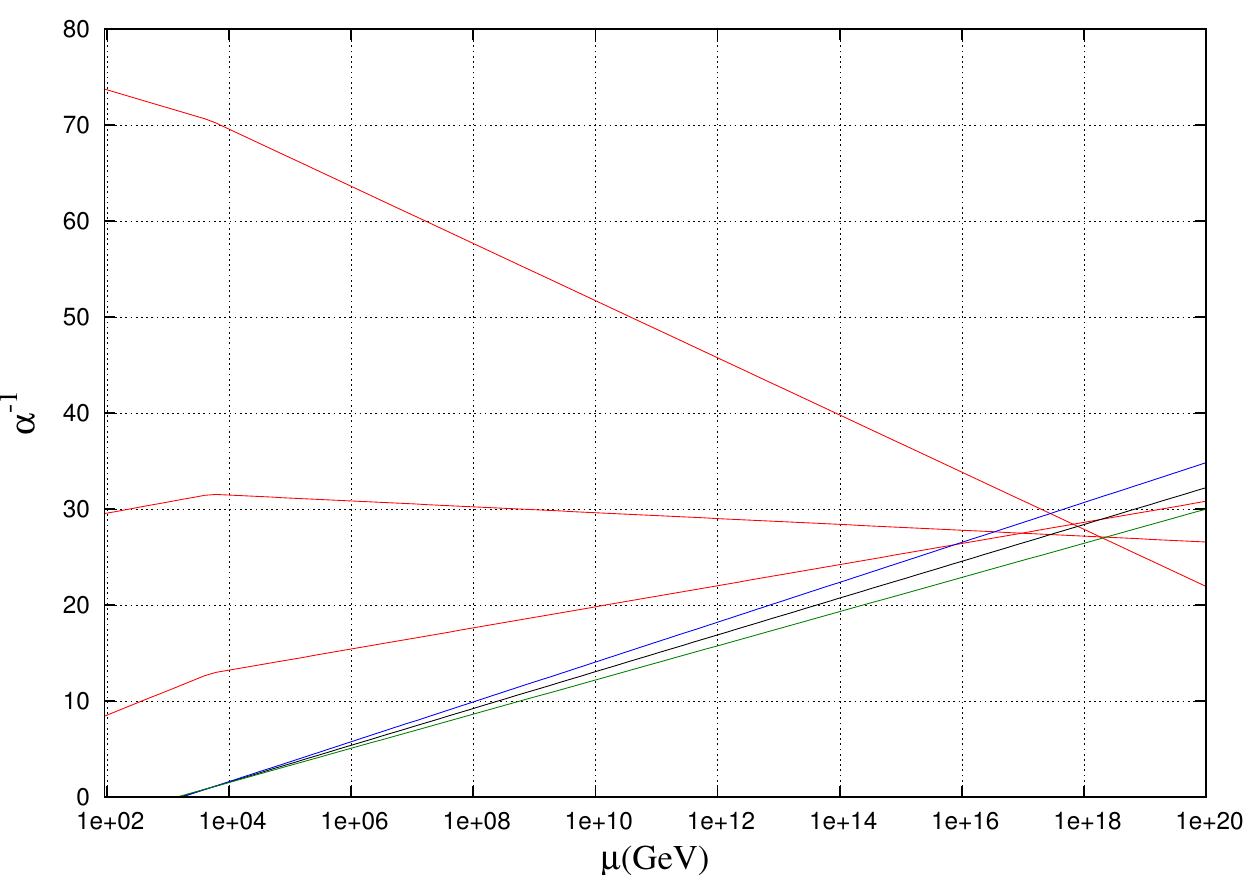}}    
\caption{\small The 1-loop running of electro-weak strong and superstrong couplings in the BSMM with the hyperchage assignment of Table~\ref{tab:alternative} and $N_S=4$ (blue line) $N_S=5$ (black line), $N_S=6$ (green line).}
\label{fig:fig7} 
\end{figure}

The presence of $N_S$ extra particles with purely superstrong vector interactions modifies the last formulae in eqs.~(\ref{NORBSMM}) and~(\ref{NORGUTBSMM}) that become 
\begin{equation}
 \mbox{Tr}\left[\left(\frac{1}{2}g_T\lambda^{3}_T\right)^2 \right]=[2(\nu_L+\nu_QN_c)+N_S]g_T^2 \, ,\label{NSGUT}
\end{equation}
and (by setting $N_c=N_T=n_g=3$ as well as $\nu_Q=\nu_L=1$)
\begin{equation}
g_4^2=\left(\frac{8+N_S}{12}\right)g_T^2 \, .\label{NSG4}
\end{equation}
This implies a modification of $ \beta_{g_4}$ in eq.~(\ref{BBSMM}) that now reads 
\begin{equation}
 \beta_{g_4}=-\frac{17}{3}\left(\frac{12}{8+N_S}\right)\frac{g^3_4}{(4\pi)^2}\, . \label{NSGB4}
\end{equation}
It is remarkable that (approximate) unification of all the four couplings can be achieved with reasonable values of $N_S$ (in the range between 4 and 6) and the natural choice for the  initial condition of $\alpha_4$ 
\beq
\alpha_4^{-1}({\mu=5\,\mbox{TeV}})=1 \, .
\label{INIT4}
\eeq
The situation is displayed in fig.~\ref{fig:fig7} where the cases $N_S=4$ (blue line), $N_S=5$ (black line) and $N_S=6$ (green line) are reported. 

\section{On the model of eq.~(\ref{LMOD})}
\label{sec:MODEL}

The inspiration for the model~(\ref{LMOD}) came from the work of ref.~\cite{Frezzotti:2014wja} (see also ref.~\cite{Frezzotti:2013raa} for an earlier version of the investigation) where a non-perturbative origin for elementary particle masses, that does not rely on the Higgs mechanism, was proposed. The complete Lagrangian of the model of ref.~\cite{Frezzotti:2014wja} (and its extension including electro-weak (EW) interactions~\cite{FRNEW}) involves a scalar SU(2) field coupled to fermions via chiral breaking Yukawa and Wilson-like terms, the latter being ``irrelevant'' operators of dimension $d=6$ appearing in the Lagrangian multiplied by two powers of the inverse UV cutoff. The structure of these terms is such that the whole Lagrangian (that is formally power-counting renormalizable) enjoys an SU$_L(2)\times$U$_Y(1)$ symmetry (under which all particles transform), crucial to forbid power divergent mass contributions in perturbation theory. 

The complete model Lagrangian is not invariant, however, under chiral SU$_L(2)\times$U$_Y(1)$ transformations of fermions and electroweak bosons only, but one can tune some parameters (the Yukawa coupling and the coefficients of the Wilson-like terms) to critical values at which the symmetry of the Lagrangian under these transformations is enforced up to UV cutoff effects, thus providing a solution of the naturalness problem in the way advocated by 't Hooft~\cite{THOOFT}. In the Nambu--Goldstone phase of the model elementary particle masses result from the non-perturbative spontaneous breaking of the restored chiral symmetry triggered by the (UV cutoff remnant of the) chiral symmetry breaking terms in the critical Lagrangian. 

The elementary particle masses turn out to be proportional to the renormalization group invariant (RGI) scale of the theory times powers of the coupling constant of the strongest interactions which the particle is subjected to. This means in particular that in order to get the right order of magnitude for the {\it top} mass the RGI scale of the whole theory must be much larger than $\Lambda_{QCD}$. This is the reason why a new superstrongly interacting sector of particles ($Q$, $L$ and gauge bosons), gauge invariantly coupled to SM matter ($q$ and $\ell$) as in eq.~(\ref{LMOD}), is postulated to exist at a few TeV scale~\footnote{We might suggestively call these new degrees of freedom ``techni-particles'' with an eye to well known techni-color models of refs.~\cite{Weinberg:1979bn,Susskind:1978ms}. We refrain from doing so as the framework underlying eq.~(\ref{LMOD}) is very different from standard techni-color. One difference is the absence of non-loop suppressed FCNC~\cite{FRNEW}. Another is that, unlike what happens in standard techni-color where techni-quarks are massless particles, our superstrongly interacting particles (SIPs) have non-perturbatively generated masses of O($\Lambda_T$) times factors of the superstrong coupling constant. In the model described by the Lagrangian~(\ref{LMOD}) (with $\nu_Q=\nu_L=1$) meson-like confined states will have masses that we can estimate, on the basis of what happens in QCD, to be of the order of two or three times $\Lambda_T$. This remark is important in the light of the existing bounds on the parameter $S$~\cite{Peskin:1990zt,Peskin:1991sw} that tend to rule out standard techni-color with more than one doublet of techni-fermions. This is not so for the particle content of the model~(\ref{LMOD}) because, as we have argued above, SIP confined  states have masses definitely ``larger'' than in standard techni-color, a fact that substantially reduces their contributions to $S$.}.

\section{Conclusions}

In this short note we have shown that it is possible to build non supersymmetric models (see, as an example, eq.~(\ref{LMOD})) where the unification of couplings is realized to a level comparable to the one it is achieved in the MSSM and in any case much better that in the SM. 

The salient feature of the BSMM described by the Lagrangian~(\ref{LMOD}) is the presence of a sector of superstrongly interacting (gluon-, quark- and lepton-like) particles with a $\Lambda_T\gg\Lambda_{QCD}$  RGI scale set in the few TeV region. Superstrongly interacting fermions are endowed with a somewhat unusual hypercharge assignment implying that the confined states have the electric charge quantized in units of $e/2$. Neutral bound states of SIPs with non-zero fermion number may provide candidates for cold dark matter along the lines discussed for techni-color, see e.g.\ in~\cite{Bagnasco:1993st}.

The motivation for studying the model~(\ref{LMOD}) stems from the work of ref.~\cite{Frezzotti:2014wja} where it is conjectured that masses of elementary particles are generated in a non-perturbative way if a (small) chiral symmetry breaking seed is present in the fundamental Lagrangian. The structure of the complete basic model of refs.~\cite{Frezzotti:2014wja,FRNEW} is dictated by two key conceptual and phenomenological requirements, namely a neat solution of the ``naturalness'' problem and the correct order of magnitude of the dynamically generated {\it top} quark mass.

\end{document}